# Maintenance Support in Open Source Software Projects

Arif Raza
Department of Computer Software Engineering
National University of Sciences and Technology, NUST
Islamabad, Pakistan
arif_raza@mcs.edu.pk

Luiz Fernando Capretz
Department of Electrical & Computer Engineering
Western University
London ON, Canada
lcapretz@uwo.ca

Faheem Ahmed
Department of Computing Science
Thompson Rivers University
Kamloops BC, Canada
fahmed@tru.ca

*Abstract*—**Easy and mostly free access to the internet has resulted in the growing use of open source software (OSS). However, it is a common perception that closed proprietary software is still superior in areas such as software maintenance and management. The research model of this study establishes a relationship between maintenance issues (such as user requests and error handling) and support provided by open source software through project forums, mailing lists and trackers. To conduct this research, we have used a dataset consisting of 120 open source software projects, covering a wide range of categories. The results of the study show that project forums and mailing lists play a significant role in addressing user requests in open source software. However according to the empirical investigation, it has been explored that trackers are used as an effective medium for error reporting as well as user requests.**

*Keywords—Open Source Software (OSS); Maintenance; Project Forums; Mailing Lists; Trackers*

## I. INTRODUCTION

Open source software confers its users the right to use, inspect, modify and distribute it both in modified or unmodified form [1]. With ever increase in the use of information technology devices; open source software is gaining popularity too. Many believe that OSS acceptance level can be raised further provided OSS addresses user requirements more rigorously [2]. According to [3], software maintainability is defined as the capability of the software systems to be modified. Software maintainability in the standard ISO/IEC 9126-1 is comprised of five sub-characteristics: analyzability, changeability, stability, testability, and maintainability compliance [3].

Koponen [4] discusses defect management and version management system as an integral part of OSS maintenance process. Aberdour [5] observes that the open source software model has led to the creation of significant pieces of software, and many of these applications show levels of quality comparable to closed source software development. According to Samoladas et al. [6], OSS development has advantages over proprietary software, such as availability of source code, functionality addition and improvement through parallel contribution by other OSS developers, regular bug fixing etc. Peer review of code by large number of testers and debuggers has certainly played its part in producing much high quality low cost open software. However lack of factors such as formal documentation and technical support are the areas where proprietary software has an edge. The authors maintain software and its maintenance deteriorates over time, neither closed source nor OSS are immune from this phenomenon. They state that *"appropriate reengineering actions may be necessary for OSS systems too."* They recommend that all three, corrective, perfective as well as preventive maintenance need to be taken into account by OSS promoters.

It is a known fact that software success is significantly dependent on its pre and post release maintenance. Software maintenance cost has a huge share in its total expenditure as well. In this work, we have studied the relationship between software maintenance issues such as error reporting and user requests and the support provided by OSS resources such as project forums, mailing lists and trackers.

In the next section we are presenting the literature review that motivated this research work. Section-3 illustrates the research model and the hypotheses of this study. This section explains the research methodology and data analysis procedure. It is followed by the discussion in Section-4 that also includes the limitations of the study. Finally the paper concludes in Section-5.

## II. LITERATURE REVIEW

Raza et al. [7] realize that since OSS user community is diverse, including both novice and computer experts, their expectations and requirements are diverse as well. The authors state that, *"with the popularity of OSS among organizations as well as among common novice users, the OSS community is no longer limited to "technically adept" individuals. Hence, the requirements and expectations of OSS are not the same as they were a decade ago, when software developers were considered to be the only OSS users."*

Chen and Huang [8] observe that although problems related to software maintenance are not the same as those of software development, there is a shortage of research that could address this issue. The authors carry out empirical investigation to study the relationship between problems in software development and software process improvement. It is believed that more than 50% budget resources are spent on software maintenance. The study concludes that software maintainability can be adversely affected by problems of software development. It is recommended that to enhance the level of software maintainability, problems associated with *"analysis, design and implementation activities during the software development phase"* need to be properly handled.

Baggen et al. [9] consider source code quality as an essential parameter of software maintainability. They point out that whenever software is required to be modified; the way its source code is developed has a huge impact on this goal. The authors however observe that assessment of code quality and its quality control are two main factors that cause either success or failure of software projects. The failure is mainly due to the absence of standardized source code framework. An approach *"for code analysis and quality consulting focused on software maintainability"* has been described. *"The approach uses a standardized measurement procedure based on the ISO/IEC 9126 definition of maintainability and source code metrics."*

Crowston and Scozzi [10] investigate the coordination practices for software bug fixing in OSS development teams and observe that task sequences are mostly sequential; submit, fix and close and effort are not equally distributed among process actors; and as a result few contribute heavily to all tasks, while the majority just submit one or two bugs.

Hedberg et al. [11] observe that with the rapid increase in the non technical users of OSS, expectations related to higher software quality will grow as well. According to them, unlike the typical OSS approach, users will not be the co-developers and competent enough to locate and fix the bugs; thus the quality assurance would need to be done before the software is delivered. They stress upon the need of having empirical research dealing with usability and quality assurance in OSS. Bevan [12] also calls for the incorporation of human centered design resources to earlier stages of software life cycle.

Schach and Offutt [13] believe that OSS outperforms closed source software mainly due to involvement and input by a large number of volunteers. According to the authors, this leads to better maintainability too since OSS is scrutinized by many voluntary coordinators. However, possibility of a lethal effect on the software as a whole by amendment of a single module cannot be ruled.

According to Zhou and Davis [14], acceptance rate of OSS is significant in the areas of server and operating systems. They observe that otherwise people are still reluctant in using OSS. The acceptance level is increasing at organizational level; however, *"there are still fears and unsolved questions especially for business people and project managers."*

Nakagawa et al. [15] quote examples of Apache web server and the GNU/Linux operating system to indicate the success of OSS. They believe that OSS success has lead to a new research area. The researchers however have shown concern about OSS quality. The authors have carried out a case study related to OSS web system development, which aims at proving a direct relationship of software architecture to OSS quality. The authors propose an *"architecture refactoring activity in order to repair software architectures, aiming at improving mainly maintainability, functionality and usability of these systems."*

Yu et al. [16] study maintainability of four OSS operating systems (Linux, FreeBSD, NetBSD, and OpenBSD) by using common coupling categorization and determining the number of global variables and the number of instances of global variables in the kernel. According to the authors, *"The coupling between two modules is a measure of the degree of interaction between those modules and, hence, of the dependency between the modules."* Based on their results the authors show concern about maintainability of Linux. The authors state, *"We believe that Linux developers need to consider controlling the use of global variables in order to balance maintainability and system efficiency."*

Although OSS is being used by all categories of users including both novice and expert professionals, it is a common perception that maintenance of open source software has a great room for improvement. Since major support for OSS maintenance is provided by online forums, their effectiveness raises questions too. Considering the issues discussed in the literature review, we embark to empirically analyze the role of these forums in providing maintenance effectively.

### III. RESEARCH MODEL AND HYPOTHESES

With the increase in number of OSS users, their requirements, demands and expectations are on the rise too. Project forums, mailing lists and trackers provide platforms for OSS users and developers to communicate. Our aim is to investigate how effectively the forums play their role towards software maintenance. In this study we look for the answer to the following research question:

*RQ: How effectively OSS forums provide software maintenance?*

The main aim of the research question (RQ) in this work is to examine the relationship between maintenance issues such as user requests and error handling, and support provided by OSS project forums, mailing lists and trackers. A research model for empirical analysis is presented in Fig. 1. The model is used for analyzing the relationship between maintenance issues and the online forums of OSS projects. Project forums are generally used for open discussion and help. Mailing lists and trackers are the two major platforms where different issues such as project releases, bug reporting, feature requests, support requests, patches, documentation are discussed.

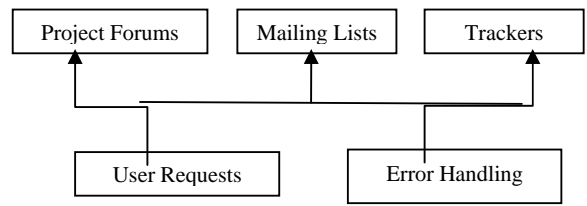

Figure -1 – Research Model

As presented in Fig. 1, the dependent variables of the model are project forums, mailing lists and trackers whereas the independent variables are user requests and error handling.

In order to empirically investigate the RQ, six hypotheses have been developed and are presented in Table I.

## A. Research Methodology

The data for this research study is collected from 120 projects of sourceforge.net which is a well-known OSS repository. This data collection is done from OSS projects of different categories including business & enterprise, communication, development, home & education, graphics, science & engineering, system administration, and security utilities. We selected top 15 projects from each category sorted on the basis of their popularity. Each of these projects has more than 500 downloads per week and is recommended by 80% of users. A filtration activity removes the data of all those projects which has no project forums, mailing list or tracker. This reduces the dataset to 92 projects. In the dataset of this study we have now business& enterprise (11), communication (10), development (11), home & education (11), graphics (14), science & engineering (14), system administration (11), and security utilities (10) projects.

TABLE I. OSS MAINTENANCE HYPOTHESES

| Hypothesis # | Statement |
| --- | --- |
| H1 | Project forums help in identifying and resolving errors in OSS projects. |
| H2 | OSS project forums assist positively in addressing user requests. |
| H3 | Error handling is positively related with mailing lists in OSS projects. |
| H4 | User requests are positively related with mailing lists in OSS projects. |
| H5 | OSS trackers help in bug tracking and error handling. |
| H6 | OSS trackers help in identifying and addressing user requests. |

The maximum project forums are found in the category of business & enterprise (40). The category of "graphics" has maximum number of mailing lists (14) in one project. The highest numbers of trackers (17) are found in a business & enterprise project.

## B. Reliability and validity of the measuring instrument

For this study we have computed the two basic parameters of an empirical study, which are reliability and validity of the measuring instrument. Reliability is about the consistency of the measurement, and validity is the power of the conjecture between the true and the measured value. For reliability analysis, internal-consistency analysis has been performed by computing coefficient alpha [17]. In our study, the coefficient alpha ranges from 0.57 to 0.90 as shown in Table II. According to van de Ven and Ferry [18], a reliability coefficient of 0.55 or higher is acceptable, whereas Osterhof [19] recommends that 0.60 or higher is satisfactory. Thus it is concluded that the measuring instruments used for this empirical investigation are reliable. A sample matrix graph between user requests and trackers is also presented in Fig. 2.

Convergent validity relates to the correlation of the scale items are and their movement in the same direction in a given work frame [20].

TABLE II. COEFFICIENT ALPHA AND PCA OF VARIABLES

| | Coefficient α | PCA Eigen value |
| --- | --- | --- |
| Project Forums – Error Handling | 0.58 | 1.07 |
| Project Forums – User Requests | 0.61 | 1.27 |
| Mailing Lists – Error Handling | 0.57 | 1.17 |
| Mailing Lists – User Requests | 0.63 | 1.25 |
| Trackers – Error Handling | 0.63 | 1.51 |
| Trackers – User Requests | 0.90 | 1.86 |

Eigen value [21] is implied as a reference point to observe the construct validity using principal component analysis (PCA).

We have used Eigen value-one-criterion, also called Kaiser Criterion ([22], [23]), which implies that any component having an Eigen value greater than one can be retained. In our Eigen value analysis, all the variables completely form a single factor, therefore the convergent validity has been considered as sufficient. It is to be noted that similar criteria has been successfully used in other empirical studies ([24], [25]), making it a proven technique.

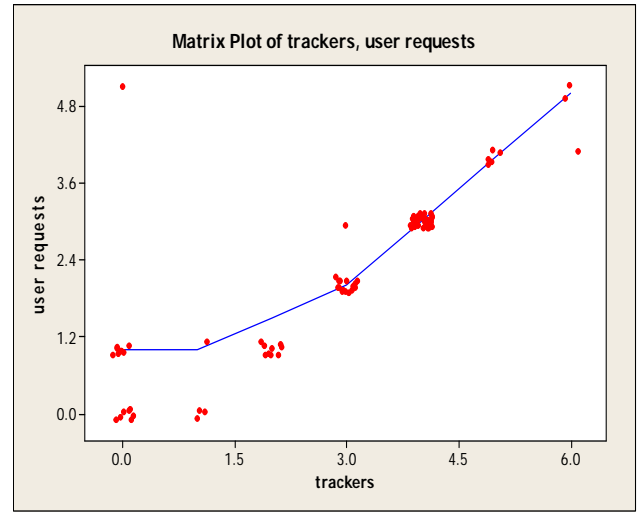

Figure-2- A Sample Matrix Graph

## C. Hypotheses analysis

To carry out the research model and analyze the hypotheses H1, H2, H3, H4, H5 and H6, statistical investigation techniques have been applied in two stages. Stage I comprises of parametric statistics for which Pearson correlation coefficients have been computed. To enhance the external validity of the study, non-parametric statistical analysis has been carried out in stage II, for which the hypotheses have been analyzed through Spearman correlation coefficients. All statistical calculations have been done using Minitab–16 Software. The results are presented in Table-III.

The Pearson correlation coefficient of 0.176 between error reporting and project forums has been found insignificant at P=0.047, thus not supporting our hypothesis H1. The hypothesis H2 has been found acceptable on the basis of the Pearson correlation coefficient (0.573) at P = 0.007, between user requests and project forums. The Pearson correlation coefficient of 0.172 at P = 0.051 has been observed indicating insignificant relation between error reporting and number of mailing lists. The hypothesis H3 has thus been rejected. The hypothesis H4 has however been found acceptable based on the Pearson correlation coefficient (0.522) at P = 0.009, between user requests and number of mailing lists. The hypothesis H5 between error reporting and trackers in OSS projects has been accepted as well considering Pearson correlation coefficient of 0.705 at P=0.000. Similarly, the hypothesis H6 between user requests and trackers in OSS projects has also been found acceptable based on Pearson correlation coefficient of 0.862 at P=0.000.

Non-parametric statistical analysis has also been carried out to increase the external validity of the study. Spearman correlation coefficients have been computed for this purpose in Stage-II, to test the hypotheses. Similar to stage I, hypothesis H1 has been found statistically insignificant at P = 0.146 with Spearman correlation coefficient of 0.154. A positive association has been observed between open source project forums and user requests with Spearman coefficient of 0.634 at P = 0.000. Similar to stage –I, our statistical outcome does not support H3, which deals with the error reporting and number of mailing lists, with Spearman correlation coefficient of 0.027 at P=0.803. The Spearman correlation of 0.532 at P = 0.008 has been observed for H4. The hypothesis H5 between error reporting and trackers in OSS projects has been found acceptable considering Spearman correlation coefficient of 0.554 at P=0.000. Similarly, the hypothesis H6 between user requests and trackers in OSS projects has also been found acceptable based on Spearman correlation coefficient of 0.919 at P=0.000.

TABLE III. EMPIRICAL ANALYSIS RESULTS

| | Pearson Correlation | | Spearman Correlation | |
|---|---|---|---|---|
| | Error Reporting | User Requests | Error Reporting | User Requests |
| Project Forums | 0.176 P=0.047** | 0.573 P=0.007* | 0.154 P=0.146** | 0.634 P=0.000* |
| Mailing Lists | 0.172 P=0.051** | 0.522 P=0.009* | 0.027 P=0.803** | 0.532 P=0.008* |
| Trackers | 0.705 P=0.000* | 0.862 P=0.000* | 0.554 P=0.000* | 0.919 P=0.000* |

* Significant at P < 0.01, ** Insignificant at P > 0.01

## IV. DISCUSSION

Software success is directly related to its user satisfaction, whether it is proprietary or OSS ([26], [27], and [28]). Capretz and Capretz [29] state that software maintenance is actually a series of activities that keeps it operational after its release. Raza et al. [7] maintain that OSS online forums do support identification and fixation of errors significantly, in particular usability related errors. However the outcome of our study does not support this claim completely. In this research we have studied the relationship of two independent variables namely, error reporting and user requests with three dependent variables, project forums, mailing lists and trackers (refer to Fig.1). Our main objective was to study that how effectively software maintenance is provided through these support forums.

First of all we collected the data from 120 OSS projects (detail is presented in Section 3). Later on we statistically analyzed the relationship between each dependent and independent variable. According to the results, our first hypothesis H1, that assumes a positive relationship between error reporting and project forums, has not been accepted. Both parametric as well as non-parametric statistical analyses do not support H1. Hence H1 has been rejected.

Second hypothesis H2 states that "OSS project forums assist positively in addressing user requests." In stage–I, we computed Pearson correlation coefficient for parametric and in stage–II, Spearman correlation coefficient has been computed for non-parametric analysis. Both the coefficients have been observed as having significant values. The hypothesis H2 has thus been accepted.

Third hypothesis studies the relationship between mailing lists and error reporting. Once again, in both the stages, correlation coefficient values have been found insignificant at P > 0.01. The hypothesis H3 has thus been rejected.

H4 states that, "User requests are positively related with mailing lists in OSS projects." The parametric as well as non-parametric analytical results support the statement of the hypothesis. The coefficient values have been observed as significant at P < 0.01. The hypothesis has thus been accepted.

H5 and H6 relate trackers with error reporting and user requests respectively. Both of these hypotheses have been strongly supported through analysis of the data. Pearson coefficient values and Spearmen coefficient values have been found significant for both the hypotheses. H5 and H6 have thus been accepted.

### A. Evaluation criterion

Empirical researches have certain limitations such as construct validity, internal validity, external validity and reliability [30], which are applicable to our study as well. Threats to external validity limit the generalization of empirical studies [31]. To address these issues, we randomly selected data from the most active OSS repository, sourceforge.net. We realize that other collaborative techniques such as blogs, wikis and t-wikis also play part in OSS maintenance; however we have focused on the roles of project forums, mailing lists and trackers.

Although empirical research is on the rise, in particular in the domain of software engineering, it has raised ethical apprehension too ([32], [33] and [34]). We did our best to stick to proven ethics and not violate any recommended principles.

Relatively small data size is one of the limitations of our work. Initially we collected data from 120 OSS projects of different categories. We selected top 15 projects from each category sorted on the basis of their popularity. However after filtering and removing projects having no project forums, mailing list or tracker, our data set got reduced to 92 projects. Although the approach adopted in this research might have some potential threat to external validity, appropriate research procedures have been followed to enhance the reliability and validity of the study.

## V. CONCLUSION

To summarize, we conclude that as far as software maintenance is considered, online forums play a significant role. User requests are supported by project forums, mailing lists and trackers. For error reporting, trackers are more effectively used in OSS environment. This study reinforces the acumen that the OSS is getting popular and that its maintenance life cycle relies on online forums. We recommend more broad based research in this area to address issues related to OSS maintenance.